\documentclass[prl,twocolumn,preprintnumbers,nofootinbib,tightenlines,superscriptaddress]{revtex4}

\usepackage{graphicx}
\usepackage{amssymb}
\usepackage{amsmath,mathrsfs,verbatim}
\usepackage{times}
\usepackage{latexsym}
\def\beq{\begin{equation}}
\def\eeq{\end{equation}}
\def\bey{\begin{eqnarray}}
\def\eey{\end{eqnarray}}

\def\kpc{\, {\rm kpc} }

\def\lsim{\mathrel{\raise.3ex\hbox{$<$\kern-.75em\lower1ex\hbox{$\sim$}}}}
\def\gsim{\mathrel{\raise.3ex\hbox{$>$\kern-.75em\lower1ex\hbox{$\sim$}}}}

\def\kms{\, {\rm km/s} }

\newcommand{\be}{\begin{equation}}
\newcommand{\ee}{\end{equation}}

\newcommand{\GeV}{{\rm ~GeV }}

\newcommand{\MeV}{{\rm ~MeV }}

\newcommand{\vmax}{V_{\rm max}}
\newcommand{\rmax}{r_{\rm max}}

\newcounter{sec}

\begin{document}

\title{Dark Matter Halos as Particle Colliders: \\
A Unified Solution to Small-Scale Structure Puzzles from Dwarfs to Clusters}

\author{Manoj Kaplinghat}
\affiliation{Department of Physics and Astronomy, University of California, Irvine, California 92697, USA}
\author{Sean Tulin}
\affiliation{Department of Physics and Astronomy, York University, Toronto, Ontario M3J 1P3, Canada}
\author{Hai-Bo Yu}
\affiliation{Department of Physics and Astronomy, University of California, Riverside, California 92521, USA}

\date{\today}

\begin{abstract}
\vspace*{.0in}

Astrophysical observations spanning dwarf galaxies to galaxy clusters indicate that dark matter (DM) halos are less dense in their central regions compared to expectations from collisionless DM N-body simulations. Using detailed fits to DM halos of galaxies and clusters, we show that self-interacting DM (SIDM) may provide a consistent solution to the DM deficit problem across all scales, even though individual systems exhibit a wide diversity in halo properties. Since the characteristic velocity of DM particles varies across these systems, we are able to measure the self-interaction cross section as a function of kinetic energy and thereby deduce the SIDM particle physics model parameters.  Our results prefer a mildly velocity-dependent cross section, from $\sigma/m \simeq 2\; {\rm cm^2/g}$ on galaxy scales to $\sigma/m \simeq 0.1\; {\rm cm^2/g}$ on cluster scales, consistent with the upper limits from merging clusters.  
Our results dramatically improve the constraints on SIDM models and may allow the masses of both DM and dark mediator particles to be measured even if the dark sector is completely hidden from the Standard Model, which we illustrate for the dark photon model.

\end{abstract}


\maketitle

\stepcounter{sec}
{\bf \Roman{sec}. Introduction.\;}  Dark matter (DM) is the dominant form of matter in galaxies and galaxy clusters, influencing the observed motions of stars and gas, as well as the lensing of light rays from distant sources.  In turn, these observables provide gravitational tracers of DM's mass distribution, which can shed light on its unknown microphysical properties.  One important question is whether DM particles undergo elastic scattering interactions with one another.
This process, known as DM self-interactions, can leave an imprint on the structure of DM halos~\cite{Spergel:1999mh}.  Self-interactions lead to heat transport between the inner and outer parts of the halo, typically causing the inner halo to become less dense.  Numerical simulations have shown that self-interacting DM (SIDM) halos have reduced central densities compared to the cold DM (CDM) paradigm, where DM is collisionless and non-interacting~\cite{Dave:2000ar,Vogelsberger:2012ku,Rocha:2012jg,Peter:2012jh,Elbert:2014bma,Fry:2015rta}.

There are long-standing anomalies from observations of dwarf galaxies that their inner regions have less DM mass compared to CDM predictions~\cite{Moore:1994yx,Flores:1994gz}.  More recent astrophysical measurements have shown this DM mass deficit problem to be ubiquitous: it appears in satellite galaxies of the Milky Way and Andromeda~\cite{BoylanKolchin:2011de,Walker:2011zu,Tollerud:2014zha}, field dwarf galaxies within the Local Group~\cite{Kirby:2014sya,Garrison-Kimmel:2014vqa}, low surface brightness (LSB) spiral galaxies~\cite{deBlok:2001fe,deBlok:2002tg,Simon:2004sr}, high surface brightness spiral galaxies~\cite{Gentile:2004tb}, and galaxy clusters~\cite{Newman:2012nw}.  Different astrophysical mechanisms have been proposed to address these issues,
including feedback from supernovae~\cite{Navarro:1996bv,Governato:2012fa} and active galactic nuclei~\cite{2012MNRAS.422.3081M}.  However, each mechanism cannot by itself solve the mass deficit problem everywhere.

The SIDM paradigm has the potential to solve the mass deficit problem in halos across all scales.  However, a quantitative comparison with observational data is lacking, in particular for spiral galaxies and clusters.   Previous studies
relied on phenomenological profiles that share qualitatively similar features as SIDM --- namely, ``cored'' profiles with reduced central densities compared to CDM --- but it is unclear how closely these mimic true SIDM profiles inferred from simulations.  It is also unknown what inferences to draw about the self-interaction cross section and whether SIDM can successfully accommodate the diversity of DM halos exhibited in nature, with core sizes from dwarf to cluster scales ranging $\sim 0.5 - 50$ kpc.  Moreover, cored profiles do not capture the gravitational effect from baryons in regions that are baryon-dominated, as in the central regions of clusters, which tends to increase the DM density~\cite{Kaplinghat:2013xca}. 

In this {\it Letter}, we address these important issues. 
In Sec.~II, we present a semi-analytical method for modeling SIDM halos~\cite{Rocha:2012jg,Kaplinghat:2013xca}.  While numerical simulations remain the standard for studies of DM structure, it is not yet feasible for them to map out SIDM parameter space because even the simplest particle models have a rich diversity of velocity-dependent cross sections~\cite{Tulin:2012wi,Tulin:2013teo}.  Our halo model provides an important bridge between simulations, astrophysical data, and particle theories for SIDM.  In Sec.~III, we fit our model to a sample of twelve dwarf and LSB galaxies and six clusters of galaxies and infer the self-interaction cross section for these systems.  We also test our method against mock data from N-body simulations of SIDM model to show that we are able to reproduce the correct input cross section.  In Sec.~IV, we discuss the role of halo concentration in explaining the observed diversity of halo profiles for DM-dominated galaxies.

The key insight resulting from our analysis is that data on different scales,
from dwarf galaxies to clusters, provide {\it complementary} handles to disentangle DM microphysics.  Since DM velocity dispersion increases with halo mass, studying halos from dwarf galaxies to galaxy clusters allows one to probe the self-interaction cross section as a function of kinetic energy 
--- similar to tuning the beam energy in a particle collider.    In Sec.~V, we illustrate this key point by applying our results to a minimal dark photon model where self-interactions are described by a Yukawa potential.  
In Sec.~VI, we summarize our main conclusions.

\stepcounter{sec}
{\bf \Roman{sec}. SIDM halo model.\;}  Scattering between DM particles is more prevalent in the halo center where the DM density is largest. It is useful to divide the halo into two regions, separated by a characteristic radius $r_1$ where the average scattering rate per particle times the halo age ($t_{\rm age}$) is equal to unity.  Thus, 
\beq \label{rateeqn}
{\rm rate \times time} \approx \frac{\langle \sigma v \rangle}{m} \rho(r_1)\,  t_{\rm age} \approx 1 \, ,
\eeq
where $\sigma$ is the scattering cross section, $m$ is the DM particle mass,  $v$ is the relative velocity between DM particles and  $\langle ... \rangle$ denotes ensemble averaging. Since we do not assume $\sigma$ to be constant in velocity, we find it more convenient to quote $\langle \sigma v \rangle/m$  rather than $\sigma/m$.  We set $t_{\rm age}=5$ and $10$ Gyr for clusters and galaxies, respectively.   Although Eq.~\eqref{rateeqn} is a dramatic simplification for time integration over the assembly history of a halo, we show by comparing to numerical simulations that it works remarkably well.

For halo radius $r > r_1$, where scattering has occurred less than once per particle on average, we expect the DM density to be close to a Navarro-Frenk-White (NFW) profile $\rho(r) = \rho_s (r/r_s)^{-1}(1+r/r_s)^{-2}$ characteristic of collisionless CDM~\cite{Navarro:1996gj}. In the halo center, for radius $r < r_1$, scattering has occurred more than once per particle.  
Here, we expect DM particles to behave like an isothermal gas satisfying the ideal gas law $p = \rho \sigma_0^2$, where $p, \rho$ are the DM pressure and mass density and $\sigma_0$ is the one-dimensional velocity dispersion.  
Since the inner halo achieves kinetic equilibrium due to DM self-interactions, the density profile can be determined by requiring hydrostatic equilibrium, $\boldsymbol{\nabla} p = - \rho \boldsymbol{\nabla}  \Phi_{\rm tot}$. Here, $\Phi_{\rm tot}$ is the total gravitational potential from DM and baryonic matter, which satisfies Poisson's equation $\boldsymbol{\nabla}^2\Phi_{\rm tot}= 4\pi G (\rho+\rho_b)$, where $G$ is Newton's constant and $\rho_b$ is the baryonic mass density.  These equations yield
\beq
\sigma_0^2 \, \boldsymbol{\nabla}^2 \ln \rho = - 4\pi G (\rho+\rho_b)\label{eq:Jeans} \, ,
\eeq
which we solve to obtain $\rho(r)$ assuming spherical symmetry.

We model the full SIDM profile by joining the isothermal and collisionless NFW profiles together at $r=r_1$:
\beq \label{eq:SIDM-model}
\rho(r) = \left\{ \begin{array}{cc} \rho_{\rm iso}(r) \, , & \;\; r < r_1 \\ \rho_{\rm NFW}(r) \, , & \;\;  r > r_1 \end{array} \right.
\eeq 
where $\rho_{\rm iso}$ is the solution to Eq.~\eqref{eq:Jeans}.  We fix the NFW parameters ($\rho_s, r_s$) by requiring that the DM density and enclosed mass for the isothermal and NFW profiles match at $r_1$.  Thus, our SIDM halo profile is specified by three parameters: the central DM density $\rho_0 \equiv \rho(0)$, velocity dispersion $\sigma_0$, and $r_1$.  Lastly, we note that this model exhibits a two-fold degeneracy in solutions for $\langle \sigma v\rangle/m$.
We keep the smaller $\langle \sigma v\rangle/m$ solutions but note that this situation may be indicative of the degeneracy between halo profiles with cores that are growing or shrinking in time~\cite{Elbert:2014bma}. 

\begin{figure}[t]
\includegraphics[scale=0.68]{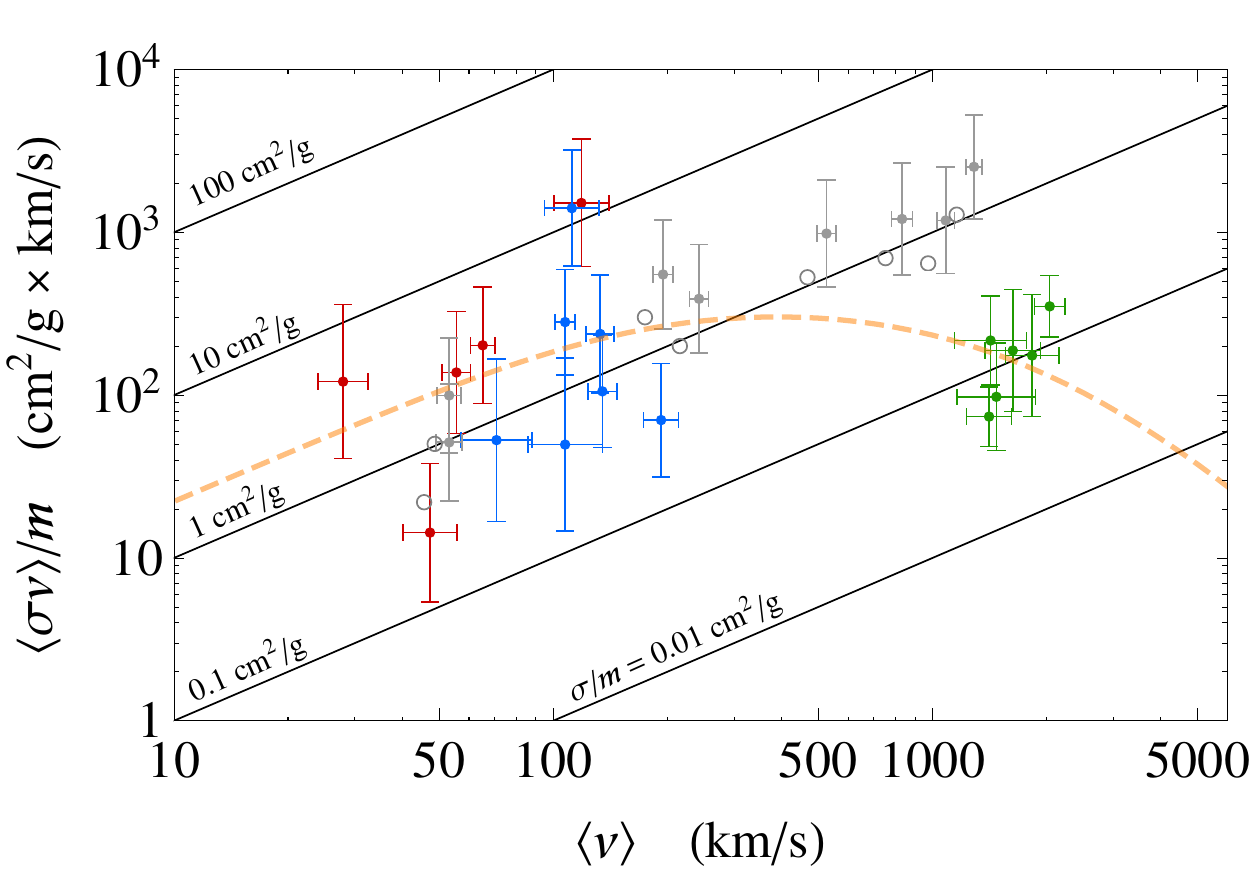} 
\caption{Self-interaction cross section measured from astrophysical data, given as the velocity-weighted cross section per unit mass as a function of mean collision velocity.  Data includes dwarfs (red), LSBs (blue) and clusters (green), as well as halos from SIDM N-body simulations with $\sigma/m=1 \; {\rm cm^2/g}$ (gray).  Diagonal lines are contours of constant $\sigma/m$ and the dashed curve is the velocity-dependent cross section from our best-fit dark photon model (Sec.~V).
}
\label{fig:sigmav}
\end{figure}


\stepcounter{sec}
{\bf \Roman{sec}. SIDM fits.\;} To constrain DM self-interactions, we consider a set of six relaxed clusters and twelve galaxies with halo masses spanning $10^{9} -  10^{15} \ M_{\odot}$.  These objects exhibit central density profiles that are systematically shallower than $\rho \propto r^{-1}$ predicted from CDM simulations.  To determine the DM profile for each system, we perform a Markov Chain Monte Carlo (MCMC) scan over the parameters $(\rho_0, \sigma_0, r_1)$ characterizing the SIDM halo, as well as the mass-to-light ratio $\Upsilon_*$ for the stellar density. 
The value for $\rho(r_1)$ determines the velocity-weighted cross section $\langle \sigma v \rangle/m$ from Eq.~\eqref{rateeqn}, as a function of average collision velocity $\langle v \rangle = (4/\sqrt{\pi}) \sigma_0$ for a Maxwellian distribution.  We also verify our model and MCMC fit procedure using a mock data set from simulations.
 
{\bf Clusters.\;} We consider the relaxed clusters from the data set of Newman, et al.~\cite{Newman:2012nw,Newman:2012nv} for which spherical modeling is appropriate (MS2137, A611, A963, A2537, A2667, and A2390). These clusters have stellar kinematics as well as strong and weak lensing measurements allowing the mass profile to be measured from stellar-dominated inner region ($\sim 10$ kpc) out to the virial radius ($\sim 3$ Mpc). The baryonic and DM densities are disentangled by constraining $\Upsilon_*$ through the assumption that all the clusters share a similar star formation history. The inferred DM density profile is consistent with CDM expectations except in the inner $\mathcal{O}(10)$ kpc region where a mass deficit is inferred~\cite{Newman:2012nw}. These small core sizes dictate the preference for a velocity-dependent cross section. 

We model each cluster using Eq.~\eqref{eq:SIDM-model} and fit directly to the stellar line-of-sight velocity dispersion data~\cite{Newman:2012nv}.  We include the gravitational effect of the stars following Eq.~\eqref{eq:Jeans} and allow for a $\pm 0.1$ dex spread in $\rho_b$ to account for systematic uncertainties~\cite{Newman:2012nw,Newman:2012nv}. Further, as a proxy for fitting to the gravitational lensing data at large radii, we fit to posteriors of the maximum circular velocity $\vmax$ and the corresponding radius $\rmax$ that have been obtained from the lensing data~\cite{Newman:2012nv}. 

\begin{figure}[t]
\includegraphics[scale=0.64]{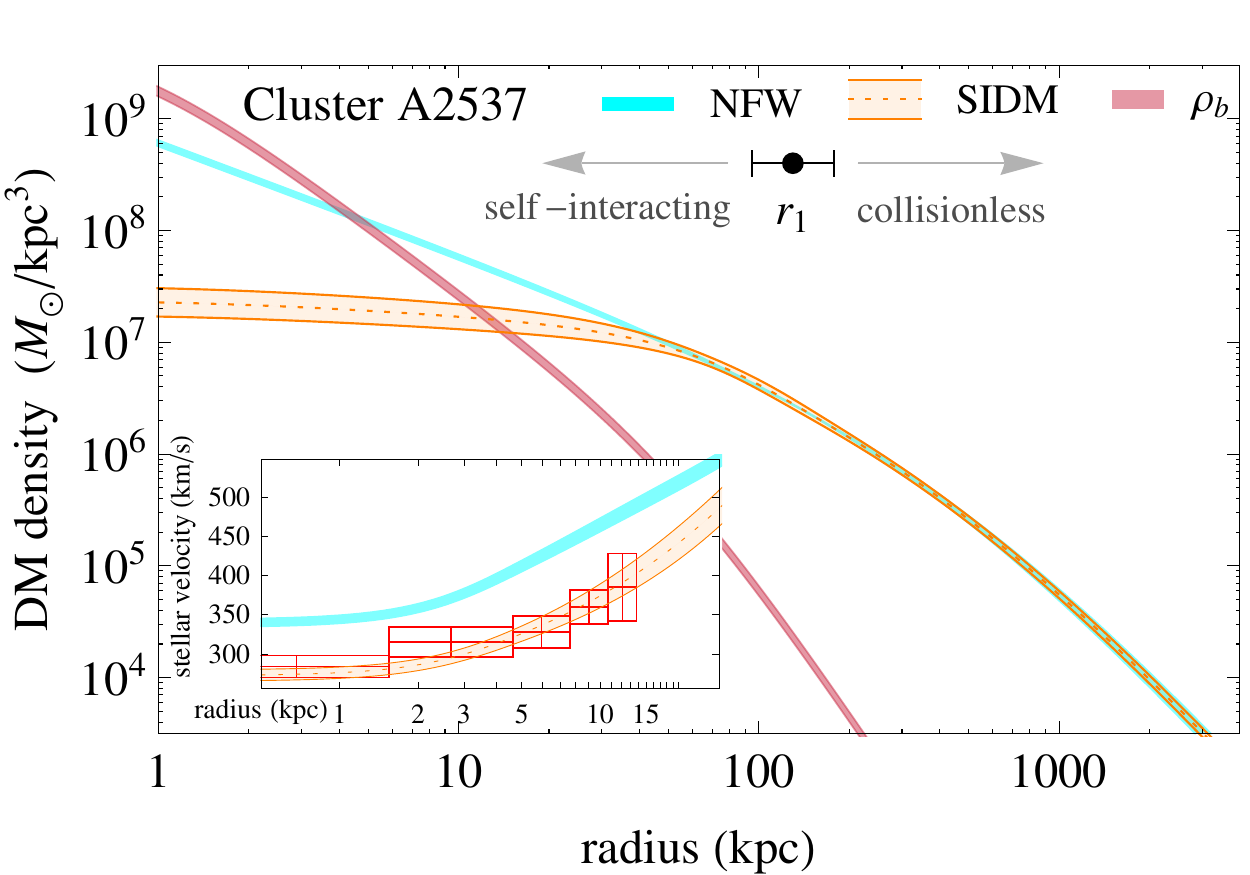} 
\includegraphics[scale=0.64]{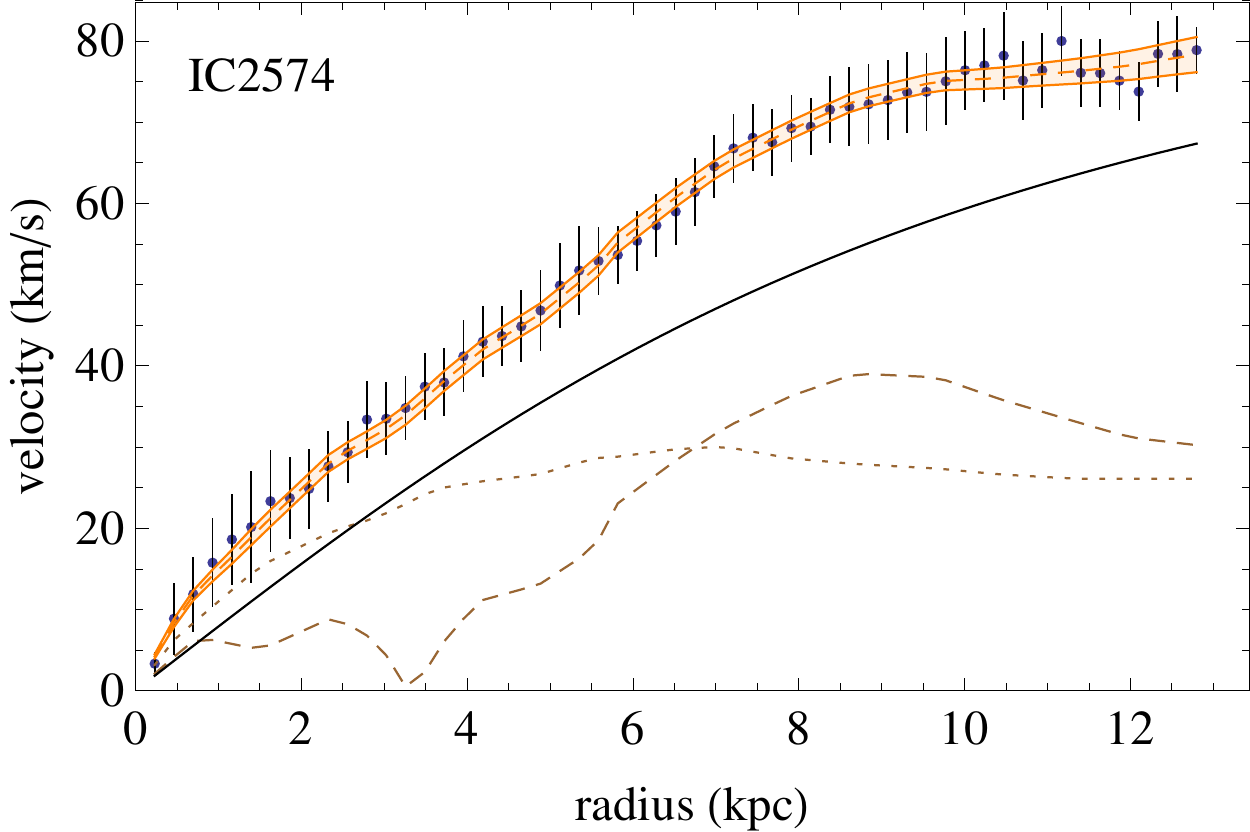} 
\caption{{\it Top:} SIDM density profile fit to cluster A2537 (orange) compared to NFW profile (cyan) and comparison to stellar kinematics data (inset). {\it Bottom:} SIDM fit to the rotation curve of galaxy IC2574 (orange) with contributions from the SIDM halo (solid), the gas disk (dashed), and stellar disk (dotted).}
\label{fig:density}
\end{figure}

As an example, we show our results for cluster A2537 in Fig.~\ref{fig:density} ({\it Top}).  Our SIDM fit is shown by the orange band ($1\sigma$ width) and the dashed line shows the mean.  The CDM prediction (cyan) is the NFW profile obtained from the gravitational lensing data~\cite{Newman:2012nv}, which provides a poor fit to the stellar kinematic data (red boxes in inset figure). The black point is the value of $r_1$ and its $1\sigma$ width.  It is reassuring that the CDM and SIDM fits, while agreeing at large radii, begin to diverge at $r_1$.  The inferred values of $\langle \sigma v \rangle/m$ for all six clusters are shown in Fig.~\ref{fig:sigmav} (green points).  Fitted with a constant cross section, we find $\sigma/m = 0.10^{+0.03}_{-0.02} \; {\rm cm^2/g}$. 

{\bf Dwarf and Low Surface Brightness Galaxies.\;} To measure DM self-interactions at small-to-intermediate scales, we consider rotation curves of five dwarf galaxies (IC 2574, NGC 2366, Ho II, M81 dwB, DDO 154) in the THINGS sample~\cite{Oh:2010ea} and seven LSB galaxies (UGC 4325, F563-V2, F563-1, F568-3, UGC 5750, F583-4, F583-1) from Kuzio de Naray, et al.~\cite{KuziodeNaray:2007qi}.  Two galaxies have been omitted from each of these samples for which $\vmax$ was not well-determined.  

To model these galaxies, we include the contributions to the rotation curve from DM, gas, and stars, with $\Upsilon_*$ allowed to vary uniformly by $\pm 0.3$ dex from the quoted population synthesis values~\cite{Oh:2010ea,deNaray:2009xj}. 
We have checked that it is a good approximation to neglect the gravitational effect of baryons on the SIDM density profile in Eq.~\eqref{eq:Jeans}.
In our likelihood, we also include a systematic error (in quadrature with the statistical error) of $5\%$ of the last measured velocity to avoid skewing our fits based on data points with small errors, $\mathcal{O}(1 \kms)$, since non-circular motions cannot be excluded at this level. 

As an example, we show the SIDM fit to the rotation curve of IC2574 in Fig.~\ref{fig:density} ({\it Bottom}).  The inferred values of $\langle \sigma v \rangle/m$ for the galaxies, shown in Fig.~\ref{fig:sigmav}, evidently prefer a larger $\sigma/m$ than the cluster measurement. Fitting all twelve galaxies with a constant cross section, we find $\sigma/m =  1.9_{-0.4}^{+0.6} \; {\rm cm^2/g}$.  We note that this value does not include systematic errors, which we discuss next.


{\bf Simulated halos.} 
To test our analytic model, we created mock rotation curve data from halos in $\sigma/m = 1\; {\rm cm^2/g}$ simulations (without baryons) and fit them with our model. Each rotation curve consisted of 20 points with a uniform 10\% velocity error and covering a range $0.1 \lesssim r/r_s \lesssim 3$ . We chose six halos with virial masses in the range $10^{11} - 10^{14} \,M_\odot$ from Ref.~\cite{Rocha:2012jg} and two dwarf-sized halos around $10^{10} \, M_\odot$ from Ref.~\cite{Elbert:2014bma}. 

The fit results shown by the gray points in Fig.~\ref{fig:sigmav} demonstrate that our simple halo model is in good agreement with results from cosmological N-body simulations for SIDM, except for the presence of a bias toward larger cross sections by a factor of $\sim 2$. 
The open circles, which also line up along $\sigma/m = 1\; {\rm cm^2/g}$, represent our SIDM profiles matched onto the ``true" NFW profile for the same halos simulated without DM self-interactions~\cite{Rocha:2012jg,Elbert:2014bma}. This analysis supports the simple picture in our model that the SIDM halo properties may be approximated by the corresponding CDM halo properties augmented with a core determined by Eq.~\eqref{rateeqn}.


\stepcounter{sec}
{\bf \Roman{sec}. Diversity.} There is considerable diversity in the properties of the galaxy cores, with almost an order of magnitude spread in density at fixed $\vmax$~\cite{deNaray:2009xj}.  This has also been recently emphasized in terms of $V_c(2\kpc)$, the measured circular velocity at 2 kpc~\cite{Oman:2015xda}, which shows a factor of $2-3$ scatter for halos with $50 \kms \lesssim \vmax \lesssim 100 \kms$.  This diversity is also reflected in the scatter in central values for $\langle \sigma v \rangle/m$ for the galaxies in Fig.~\ref{fig:sigmav}.

How does this scatter arise in our model? The answer is surprising in its simplicity: it is directly related to the halo assembly history.  
Different formation histories encoded in ($\rho_s,r_s$) values (essentially the CDM halo-to-halo scatter) lead to SIDM halos with different core sizes and central densities through Eq.~\eqref{eq:SIDM-model}. 
This explanation is implicit in Fig.~\ref{fig:sigmav} where the large errors on $\langle \sigma v \rangle$ reflect, partly, the lack of constraints on ($\rho_s,r_s$).  Choosing the ``right'' value of ($\rho_s,r_s$) for each galaxy would reduce the scatter in $\langle\sigma v\rangle/m$ considerably.
 
If we fix the $\rho_s$-$r_s$ relation to its median in $\Lambda$CDM cosmology~\cite{Prada:2011jf} in our analysis, the galaxies UGC 5750 and IC 2574 prefer the largest cross sections, $\sigma/m \sim 10 \; {\rm cm^2/g}$, while M81 dwB prefers the smallest cross sections, $\sigma/m \sim 0.1 \; {\rm cm^2/g}$. 
However, if UGC 5750 and IC 2574 halos are $2\sigma$ less concentrated and M81 dwB halo $2\sigma$ more concentrated than the median halo, the inferred $\sigma/m$ become consistent, within errors, with $\sim 1\; {\rm cm^2/g}$.

Turning this around, we can fix $\sigma/m$ and look at the impact of the scatter in the $\rho_s$-$r_s$ relation on $V_c(2\kpc)$.  Within our analytic model, we have checked that the spread in the $\rho_s$-$r_s$ relation in $\Lambda$CDM leads to about a factor of two spread in $V_c(2\kpc)$ for the relevant galaxies. If we were to add baryons (which could be important within 2 kpc), it is conceivable that the bulk of the spread seen in Ref.~\cite{Oman:2015xda} can be explained. 


\stepcounter{sec}
{\bf \Roman{sec}. Dark matter particle properties.} The energy dependence of the cross section allows one to discern the underlying particle dynamics of SIDM.  
The data in Fig.~\ref{fig:sigmav} range over a factor of $10^4$ in kinetic energy and prefer a cross section that mildly falls with energy. 

To illustrate the implications for particle physics, let us consider the dark photon model for DM self-interactions.  In this model, self-interactions are governed by a Yukawa potential, $V(r) = \alpha^\prime e^{- \mu r}/r$, 
where $\alpha^\prime$ is the coupling constant (analogous to the fine structure constant $\alpha \approx 1/137$) and $\mu$ is the dark photon mass, which screens the potential \cite{Feng:2009hw,Buckley:2009in,Loeb:2010gj}. To be concrete, we will set $\alpha^\prime = \alpha$.  We then compute $\langle \sigma v \rangle/m$ using standard partial wave methods discussed in Ref.~\cite{Tulin:2013teo}.  
Comparing the theoretical predictions to the data points in Fig.~\ref{fig:sigmav} using a $\Delta \chi^2$ test, we determine the preferred regions for the DM mass $m$ and dark photon mass $\mu$ .  
To take into account the uncertainty in our modeling (apparent in our predictions for the simulated halos), we have included an additional systematic uncertainty (in quadrature) of $\Delta(\log \langle \sigma v \rangle/m) = 0.3$ and $\Delta(\log \langle v \rangle ) = 0.1$ for each system.

\begin{figure}[t]
\includegraphics[scale=0.63]{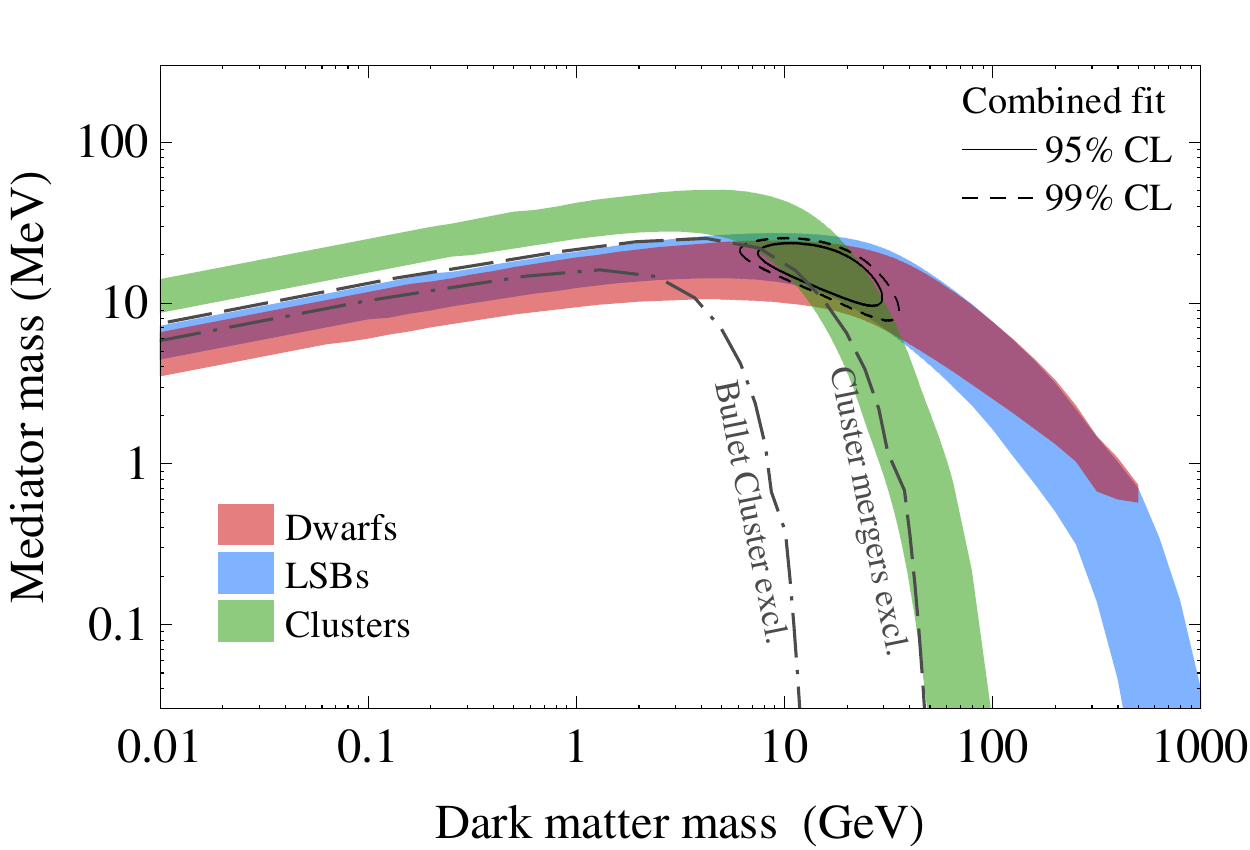} 
\caption{Parameter space for dark photon model of self-interactions (with $\alpha^\prime = \alpha$), preferred by dwarfs (red), LSB spiral galaxies (blue), and clusters (green), each at 95\% CL.  Combined 95\% (99\%) region is shown by the solid (dashed) contours.  The estimated Bullet Cluster excluded region lies below dot-dashed curve and the ensemble merging cluster excluded region below the long-dashed curve.
}
\label{fig:plot2}
\end{figure}

Our results shown in Fig.~\ref{fig:plot2} illustrate the important complementarity between observations across different scales in constraining DM microphysics.  The red, blue, and green shaded bands show the individual 95\% confidence level (CL) regions preferred by our analysis of dwarf galaxies, LSBs, and clusters, respectively. The solid (dashed) black contour shows the 95\% (99\%) CL region from all observations combined. These data prefer DM mass of $15^{+7}_{-5} \GeV$ and dark photon mass of $17 \pm 4 \MeV$ at 95\% CL. For the best-fit values of $m$ and $\mu$, we plot $\left<\sigma v\right>/m$ as a function of $\left<v\right>$ in Fig.~\ref{fig:sigmav} (dashed).

Fig.~\ref{fig:plot2} also shows the regions excluded by the Bullet Cluster constraint of $\sigma/m < 1.25 \; {\rm cm^2/g}$ at 68\% CL~\cite{Randall:2007ph} at $v=2000 \kms$ (dot-dashed) and the constraint from an ensemble of merging clusters of $\sigma/m<0.47\; {\rm cm^2/g}$ at 95\% CL~\cite{Harvey:2015hha} at  $v=900 \kms$ (long-dashed). 
A more refined analysis of the merging clusters, including large dissociative clusters that show offsets between the luminous and dark components~\cite{Randall:2007ph,Dawson:2011kf,Jee:2014hja,Jee:2014mca}, would be interesting in light of the velocity dependence.

It is remarkable that astrophysical observations can pick out a closed range for the DM mass $m$ (albeit within the simple model we have adopted).  For $m \lesssim 10~{\rm GeV}$, the cross section changes little with velocity, $\sigma \propto m^2/\mu^4$, which is disfavored by the velocity dependence evident in Fig.~\ref{fig:sigmav}.  For $m \gtrsim 100~\GeV$, cross section tends to the Rutherford limit, $\sigma \propto 1/(m^2 v^4)$, which is too steep a velocity dependence to be consistent with our fits. The preferred region lies in between these extremes: $\sigma$ is constant at small velocity and turns over to a Rutherford-like dependence at large velocity.

\stepcounter{sec}
{\bf \Roman{sec}. Conclusions.\;} SIDM paradigm may provide a unified explanation for the apparent deficit of DM in the central regions of galaxies and clusters.  We have explored the direct connection between self-interactions and astrophysical observations for a set of twelve galaxies and six clusters using a simple model for SIDM halos calibrated to N-body simulations.  Despite the diversity of DM halo properties in these systems, the majority of dwarfs and LSBs is remarkably consistent with $\sigma/m\approx 2~{\rm cm^2/s}$. Clusters favor $0.1~{\rm cm^2/g}$ because their halo profiles are largely consistent with CDM except in the inner $\mathcal{O}(10 \kpc)$ region. The velocity dependence discernible in these data provides an important step toward understanding the possible particle physics of DM self-interactions. Within the dark photon model we considered, these data prefer DM mass of $\sim15\GeV$ and dark photon mass of $\sim17\MeV$. While these conclusions are model-specific, SIDM in general indicates a new mass scale much below than the electroweak scale. Using DM halos as particle colliders, we may be able to unveil the particle physics nature of DM, independent of whether the dark and visible sectors are coupled via interactions beyond gravity.  

{\it Acknowledgments}:  We thank A.~Newman for helpful discussions, O.~Elbert for making available his simulation data, and S.-H. Oh and S. McGaugh for providing the rotation curve data.  This work was supported by the National Science Foundation Grant No.~PHY-1214648 and PHY-1316792 (MK), the Natural Science and Engineering Research Council of Canada (ST), the U. S. Department of Energy under Grant No.~DE-SC0008541 (HBY), as well as by the National Science Foundation under Grant No.~PHY11-25915 as part of the KITP ``Particle-genesis'' workshop (ST \& HBY).

\bibliography{clustercore}

\end{document}